\RequirePackage{fix-cm}
\documentclass[smallextended]{svjour3}       % onecolumn (second format)
\smartqed  % flush right qed marks, e.g. at end of proof
\usepackage{graphicx}
\usepackage{amsmath}
\usepackage{amssymb}
\usepackage{color}
\newcommand{\QED}{\quad \hfill {$\square$}}
\newcommand{\define}{\stackrel{\rm def.}{=}}
\renewcommand{\mod}{\mbox{\rm mod}\;}

\begin{document}

\title{A 2.75-Approximation Algorithm for the 
       Unconstrained Traveling Tournament Problem\thanks{%
The present study was supported in part by Grants-in-Aid for Scientific 
Research, by the Ministry of Education, Culture, Sports, Science 
and Technology of Japan.}
}

% \subtitle{Do you have a subtitle?\\ If so, write it here}

\titlerunning{A 2.75-Approximation Algorithm for the Unconstrained TTP}% if too long for running head

\author{
\mbox{
Shinji Imahori \and    
Tomomi Matsui  \and    
Ryuhei Miyashiro
}
}

\authorrunning{S.~Imahori, T.~Matsui, R.~Miyashiro}% if too long for running head

\institute{
S.~Imahori \at
 Graduate School of Engineering, Nagoya University,\\
 Furo-cho, Chikusa-ku, Nagoya 464-8603, Japan.\\
 \email{imahori@na.cse.nagoya-u.ac.jp}
 \and
T.~Matsui \at
 Faculty of Science and Engineering, Chuo University, \\
 Kasuga, Bunkyo-ku, Tokyo 112-8551, Japan. \\
 \email{matsui@ise.chuo-u.ac.jp}
 \and
R.~Miyashiro \at
 Institute of Engineering,
 Tokyo University of Agriculture and Technology,\\
 Naka-cho, Koganei, Tokyo 184-8588, Japan. \\
 \email{r-miya@cc.tuat.ac.jp}
}

\date{Received: date / Accepted: date}

\maketitle

\begin{abstract}
A 2.75-approximation algorithm is proposed 
	for the unconstrained traveling tournament problem, 
	which is a variant of the traveling tournament problem.
For the unconstrained traveling tournament problem, 
	this is the first proposal of an approximation algorithm 
	with a constant approximation ratio. 
In addition, the proposed algorithm yields a solution
	that meets both the no-repeater and mirrored constraints.
Computational experiments show that the algorithm generates solutions of good quality.

\end{abstract}

\section{Introduction}

In the field of tournament timetabling, the traveling tournament problem (TTP) 
	is a well-known benchmark problem 
	established by Easton, Nemhauser, and Trick~\cite{Easton2001}.
The present paper considers the unconstrained traveling tournament problem (UTTP),
	which is a variant of the TTP\@.
In the following, some terminology and the TTP are introduced.
The UTTP is then defined at the end of this section.

Given a set~$T$ of $n$ teams, where $n \geq 4$ and is even,
	a game is specified by an ordered pair of teams. 
Each team in~$T$ has its home venue. 
A double round-robin tournament is a set of games 
	in which every team plays every other team once at its home venue 
	and once as an away game (i.e., a game held at the home venue of the opponent).
Consequently, $2(n-1)$ slots
	are necessary to complete a double round-robin tournament. 

Each team stays at its home venue before a tournament 
	and then travels to play its games at the chosen venues. 
After a tournament, 
	each team returns to its home venue if the last game is played as an away game.  
When a team plays two consecutive away games, 
	the team goes directly from the venue of the first opponent 
	to the venue of another opponent without returning to its home venue. 

For any pair of teams $i, j \in T$, $d_{ij} \geq 0$ denotes 
	the distance between the home venues of $i$ and~$j$.
Throughout the present paper, we assume that 
	triangle inequality ($d_{ij} + d_{jk} \geq d_{ik}$), 
	symmetry ($d_{ij} = d_{ji}$), and 
	$d_{ii} = 0$ hold for any teams~$i, j, k \in T$.

Denote the distance matrix $(d_{ij})$ by $D$.
Given an integer parameter $u \geq 2$, 
	the traveling tournament problem~\cite{Easton2001} is defined as follows. 

\medskip
\noindent
{\bf Traveling Tournament Problem (TTP$(u)$)}\\
{\bf Input:\/} A set of teams $T$ and 
		a distance matrix~$D=(d_{ij})$. \\
{\bf Output:\/}
  A double round-robin schedule of $n$~teams such that

\noindent
C1. No team plays more than $u$ consecutive away games,

\noindent
C2. No team plays more than $u$ consecutive home games, 

\noindent
C3. Game $i$ at~$j$ immediately followed by game $j$ at~$i$ 
       is prohibited, 

\noindent
C4. The total distance traveled by the teams is minimized. 

\medskip 

\noindent
Constraints C1 and~C2 are referred to as the {\em atmost\/} constraints,
	and Constraint~C3 is referred to as the {\em no-repeater\/} constraint.

\medskip

Various studies on the TTP have been conducted in recent years 
	(see \cite{Kendall2010,Rasmussen2006-2,Trick2011} for detail), 
	and most of these studies considered TTP(3)~\cite{Trick_web},
	which was recently proved to be NP-hard by Thielen and Westphal~\cite{NPhard}.
Almost all of the best upper bounds of TTP instances are obtained 
	using metaheuristic algorithms. 
On the other hand, 
	little research on approximation algorithms has been conducted for the TTP\@. 
Miyashiro, Matsui, and Imahori~\cite{MMI} proposed 
	a $(2+O(1/n))$-approximation algorithm for TTP(3).
Yamaguchi, Imahori, Miyashiro, and Matsui~\cite{YMIM} proposed
	an approximation algorithm for TTP$(u)$, where $3 \le u \ll n$.
Westphal and Noparlik~\cite{Westphal2010} proposed\footnote{
	Westphal and Noparlik's paper~\cite{Westphal2010} and the conference version 
	of the present paper~\cite{IMM} appeared in the same conference 
	(PATAT, 2010).
}
	a 5.875-approximation algorithm for TTP$(u)$, where $3 \le u$.
For TTP(3), the approximation ratio of~\cite{YMIM} is the best among them.
In addition, Thielen and Westphal~\cite{Thielen2010} proposed 
	a $(1.5+O(1/n))$-approximation algorithm for TTP(2).

The TTP is a simplification of an actual sports scheduling problem. 
Some further simplified variants of the TTP have been studied~\cite{Trick_web}. 
The circular distance TTP and the constant distance TTP are the problems 
which have specific distance matrices. 
For the constant distance TTP, 
Fujiwara, Imahori, Matsui, and Miyashiro~\cite{Fujiwara2007}
proposed approximation algorithms. 

The unconstrained traveling tournament problem (UTTP) is another variant of the TTP,
	in which Constraints C1 through~C3 are eliminated.
% The unconstrained traveling tournament problem (UTTP) is a variant of the TTP,
% 	in which Constraints C1 through~C3 are eliminated.
In other words, 
	the UTTP is equivalent to TTP($n-1$) without the no-repeater constraint.
On some actual sports scheduling problems, the atmost constraints 
	($u=3$ in particular) 
	and the no-repeater constraint are considered. 
However, these constraints are not necessarily imposed,  
	and the UTTP is a suitable simplified model for some practical scheduling problems.  

Bhattacharyya~\cite{NPC} recently showed NP-hardness of the UTTP\@.
Although the UTTP is simpler than the TTP,
	no approximation algorithm has yet been proposed for the UTTP\@.
The method proposed in~\cite{YMIM} cannot be applied to the UTTP
	because the condition $u \ll n$ is necessary.
The method in~\cite{MMI}, proposed for TTP(3), 
	can be applied to the UTTP with a few modifications. 
However, this leads to a $((2/3)n+ O(1))$-approximation algorithm for the UTTP, 
	which is not a constant approximation ratio with regard to~$n$.

In the present paper, 
	we propose a 2.75-approximation algorithm for the UTTP\@.
In addition, the solution obtained by the algorithm meets
	both the no-repeater and mirrored constraints,
	which are sometimes required in practice.
This property indicates that our algorithm also works for TTP($n-1$), 
which eliminates the atmost constraints but considers the no-repeater constraint.

%%%%%%%%%%%%%%%%%%%%%%%%%%%%%%%%%%%%%%%%%%%%%%%%%%%%%%%%%%%%%%%%%%%%%%
\section{Algorithm}

In this section, we propose an approximation algorithm for the UTTP\@.
A key concept of the algorithm is the use of the circle method and a shortest Hamilton cycle.
The classical schedule obtained by the circle method 
	satisfies the property that, for all teams but one,
	the orders of opponents
	are very similar to a mutual cyclic order.
Roughly speaking, the proposed algorithm constructs 
	a short Hamilton cycle passing all venues, 
	and finds a permutation of teams 
	such that the above cyclic order corresponds 
	to the Hamilton cycle.

Let $G=(V,E)$ be a complete undirected graph with the vertex set~$V$ and edge set $E$, where $|V|=n$.
We assume that there exists a bijection between the vertex set $V$ and the set of teams $T$.
We put the length of edge $\{v,v'\} \in E$, denoted by  $d_{vv'}$, 
	to the distance between the home venues of the corresponding teams $t,t' \in T$.
First, we assign aliases $0,1,\ldots,n-1$ to teams in $T$ as follows.
\begin{enumerate}
\item For each $v \in V$, compute $\sum_{v' \in V \setminus\{v\}} d_{vv'}$.
\item Let $v^*$ be a vertex 
	that attains $\min_{\, v \in V} \sum_{v' \in V \setminus\{v\}} d_{vv'}$,
	and designate the team corresponding to $v^*$ as team $n-1$.
\item
	Using Christofides' 1.5-approximation algorithm~\cite{Christofides} 
		for the traveling salesman problem with triangle inequality and symmetry,
		construct a Hamilton cycle on the complete graph induced by $V \setminus \{v^*\}$.
	For the obtained cycle $(v_{0},v_{1},\ldots,v_{n-2})$,
		denote the corresponding teams by $(0,1,\ldots ,n-2)$.
\end{enumerate}

\noindent
In the rest of this paper, we define that 
	the set of teams $T=\{0,1,2,\ldots,n-1\}$ and 
	the vertex set $V=\{v_{0}, v_{1},\ldots, v_{n-2}, v^*\}$.
We identify the vertex $v_{n-1}$ with~$v_0$ (not $v^*$) and
	the vertex~$v_{-1}$ with $v_{n-2}$ (not $v^*$). 

Next, we construct a single round-robin schedule.
In the following, a ``schedule without HA-assignment'' refers to a
	``round-robin schedule without the concepts of home game, away game, and venue.''
Denote the set of $n-1$ slots by $S= \{ 0, 1, \ldots, n-2 \}$.
A single round-robin schedule 
	without HA-assignment is a matrix~$K$ of which $(t,s) \in T \times S$ element, 
	say $K(t,s)$, denotes the opponent of team~$t$ in slot~$s$.
Let $K^*$ be a matrix defined by
\[
	K^*(t,s)= \left\{ \begin{array}{ll}
		s-t \; (\mod n-1) 	& (t \neq n-1 \mbox{ and } s-t \neq t \ (\mod  n-1)), \\
		n-1					& (t \neq n-1 \mbox{ and } s-t = t \ (\mod n-1)), \\
		s/2					& (t = n-1    \mbox{ and } s \mbox{ is even}), \\
        (s+n-1)/2			& (t = n-1    \mbox{ and } s \mbox{ is odd}). \\
	\end{array} \right.
\]
\begin{lemma}{\rm{\cite{YMIM}}} \label{lem1}
	The matrix $K^*$ is a single round-robin schedule without HA-assignment.
In addition, $K^*$ is essentially equivalent to the classical schedule obtained by the circle method.
\end{lemma}

Then, by the mirroring procedure, 
	we form $K^*$ into a double round-robin schedule without HA-assignment.
More precisely, construct a matrix~$(K^* | K^*)$
	whose rows are index by teams 
	and columns are index by a sequence of slots $(0,1,\ldots,n-2,n-1,n,\ldots,2n-3)$.
So as to complete a double round-robin schedule, ``home'' and ``away'' are assigned to games
of $(K^* | K^*)$ as follows:
\begin{itemize}
% \item for team $t \in \{0, 1, \ldots, n/2-1 \}$,
%       let the games in slots $n+2t-1, n+2t,\ldots, n+2t+n-3 \ (\mod 2n-2)$
% 	be away games, and let the other games be home games.
\item for team $t \in \{0, 1, \ldots, n/2-1 \}$,
      let the games in slots $2t, 2t+1,\ldots, n+2t-2$
	be home games, and let the other games be away games.
\item for team $t \in \{n/2, n/2+1, \ldots, n-2 \}$,
      let the games in slots $2t-n+2, 2t-n+3, \ldots, 2t$ be away games,
      and let the other games be home games.
\item for team $n-1$, let the games in slots $0,1,\ldots,n-2$ be away games,
      and let the other games be home games.
\end{itemize}
The obtained double round-robin schedule is denoted by $K^*_{\mathrm{DRR}}$.
Figure~\ref{KDRR} shows the schedule $K^*_{\mathrm{DRR}}$ of 10 teams.
\begin{figure}[htb]
\begin{center}
{\small
\noindent
\begin{tabular}{r|c@{ \,\,}c@{ \,\,}c@{ \,\,}c@{ \,\,}c@{ \,\,}c@{ \,\,}c@{ \,\,}c@{ \,\,}c@{ \,\,}c@{ \,\,}c@{ \,\,}c@{ \,\,}c@{ \,\,}c@{ \,\,}c@{ \,\,}c@{ \,\,}c@{ \,\,}c}
 slots \\
 teams \,\,\,& 0 & 1 & 2 & 3 & 4 & 5 & 6 & 7 & 8 & 9 & 10&11 &12 &13 &14 &15 &16 &17 \\
 \hline
0 & 9H& 1H& 2H& 3H& 4H& 5H& 6H& 7H& 8H& 9A& 1A& 2A& 3A& 4A& 5A& 6A& 7A& 8A \\
1 & 8A& 0A& 9H& 2H& 3H& 4H& 5H& 6H& 7H& 8H& 0H& 9A& 2A& 3A& 4A& 5A& 6A& 7A \\
2 & 7A& 8A& 0A& 1A& 9H& 3H& 4H& 5H& 6H& 7H& 8H& 0H& 1H& 9A& 3A& 4A& 5A& 6A \\
3 & 6A& 7A& 8A& 0A& 1A& 2A& 9H& 4H& 5H& 6H& 7H& 8H& 0H& 1H& 2H& 9A& 4A& 5A \\
4 & 5A& 6A& 7A& 8A& 0A& 1A& 2A& 3A& 9H& 5H& 6H& 7H& 8H& 0H& 1H& 2H& 3H& 9A \\
5 & 4H& 9H& 6A& 7A& 8A& 0A& 1A& 2A& 3A& 4A& 9A& 6H& 7H& 8H& 0H& 1H& 2H& 3H \\
6 & 3H& 4H& 5H& 9H& 7A& 8A& 0A& 1A& 2A& 3A& 4A& 5A& 9A& 7H& 8H& 0H& 1H& 2H \\
7 & 2H& 3H& 4H& 5H& 6H& 9H& 8A& 0A& 1A& 2A& 3A& 4A& 5A& 6A& 9A& 8H& 0H& 1H \\
8 & 1H& 2H& 3H& 4H& 5H& 6H& 7H& 9H& 0A& 1A& 2A& 3A& 4A& 5A& 6A& 7A& 9A& 0H \\
9 & 0A& 5A& 1A& 6A& 2A& 7A& 3A& 8A& 4A& 0H& 5H& 1H& 6H& 2H& 7H& 3H& 8H& 4H \\
\end{tabular}
}
Each number corresponds to the opponent and away (home) game is denoted by A (H). 
\caption{The schedule $K^*_{\mathrm{DRR}}$ with 10 teams. } \label{KDRR}
\end{center}
\end{figure}
\begin{lemma} \label{newlem}
	The double round-robin schedule $K^*_{\mathrm{DRR}}$ is feasible.

	{\rm \noindent
{\bf Proof.\/}
$(K^* | K^*)$ is a consistent double round-robin schedule without HA-assignment, 
	which satisfies the mirrored constraint. 
We check the feasibility of HA-assignment to games. 
Teams $i$ and $j$ ($i < j < n-1$) have a game at slot~$i+j$. 
By the rule to assign home and away to games, 
	team~$i$ plays a home game and team~$j$ plays an away game at slot~$i+j$. 
Teams $i$ and $j$ ($i < j = n-1$) have a game at slot~$2i$,  
	and the rule assigns consistent home/away to the teams. 
Another game between teams~$i$ and $j$ is held at the opposite venue. \QED
}
\end{lemma}
In addition, for each $m \in \{0,1,\ldots ,2n-3\}$,
	we construct a double round-robin schedule by
	rotating slots of $K^*_{\mathrm{DRR}}$ through $m$ cyclically.
It means that games of $K^*_{\mathrm{DRR}}(m) \ (m \in \{0,1,\ldots ,2n-3\})$
	at slot~$s$ are equal
	to games of $K^*_{\mathrm{DRR}}$ at slot~$s+m \ (\mod 2n-2)$.
Obviously, all of the schedules $K^*_{\mathrm{DRR}}(m) \ (m \in \{0,1,\ldots ,2n-3\})$ meet
	both the no-repeater and mirrored constraints.
Finally, output a best solution among~$K^*_{\mathrm{DRR}}(m)$ $ (m \in \{0,1,\ldots ,2n-3\})$.

Here, we estimate the time complexity of the algorithm. 
Christofides' algorithm requires $O(n^3)$ time 
	to construct a Hamilton cycle on the complete graph induced by $V \setminus \{v^*\}$. 
For the constructed Hamilton cycle, there are $2(n-1)$ possibilities to assign teams.   
For each assignment of teams, we consider $2n-2$ possibilities of $m \in \{0,1,\ldots ,2n-3\}$. 
Each double round-robin schedule can be evaluated in $O(n)$ time on average. 
Thus, the time complexity of the algorithm is bounded by $O(n^3)$. 

In the next section, we prove that the proposed algorithm guarantees an approximation ratio~2.75.
% \medskip

%%%%%%%%%%%%%%%%%%%%%%%%%%%%%%%%%%%%%%%%%%%%%%%
\section{Approximation Ratio}

In this section, 
	we describe the proof of the approximation ratio of the proposed algorithm.
Designate the length of a shortest Hamilton cycle on~$G$ as $\tau$. 

\begin{lemma}\label{lem2}
The following propositions hold for $G$. 

{\rm (1)}  The length of any edge is bounded by $\tau /2$.

{\rm (2)}  The length of any Hamilton cycle on $G$ is bounded by $n\tau /2$.

{\rm (3)} $\displaystyle  
	\sum_{v \in V} \sum_{v' \in V \setminus \{v\}} d_{vv'} \leq n^2 \tau /4$.

{\rm (4)} $\displaystyle  
	\sum_{v \in V \setminus \{v^*\}}d_{vv^*} \leq n \tau /4$. 

{\rm \noindent
{\bf Proof.\/}	
(1) 
For the edges $\{i,j\}$ and~$\{j,i\}$, the sum of their lengths
	is at most the length of a shortest Hamilton cycle.
Thus, the length of the edge~$\{i,j\}$ is bounded by $\tau /2$ with symmetry.
	
\noindent
(2) 
This immediately follows from Property~(1).

\noindent
(3) 
Given a shortest Hamilton cycle $H=(u_0,u_1,\ldots,u_{n-1})$ on $G$, let
\[
	h_{u_i, u_j} = \left\{ \begin{array}{ll}
		d_{u_i, u_{i+1}} + d_{u_{i+1}, u_{i+2}} + \cdots + d_{u_{j-1}, u_j}	& (j - i \ (\mod  n) \le n/2), \\
		d_{u_i, u_{i-1}} + d_{u_{i-1}, u_{i-2}} + \cdots + d_{u_{j+1}, u_j}	& (j - i \ (\mod  n) > n/2). \\
	\end{array} \right.
\]
Then, we have:
\begin{eqnarray*}
	\sum_{v \in V} \sum_{v' \in V \setminus \{v\}} d_{vv'} &=&
	\sum_{i=0}^{n-1} \sum_{k=1}^{n-1} d_{u_i, u_{i+k \, ({\rm mod} \, n)}} \\
	&\leq& 	\sum_{k=1}^{n-1} \sum_{i=0}^{n-1} h_{u_i, u_{i+k \, ({\rm mod} \, n)}} \\
	&=&  \sum_{k=1}^{n/2-1} \sum_{i=0}^{n-1} \left( d_{u_i, u_{i+1}} + d_{u_{i+1}, u_{i+2}} + \cdots + d_{u_{i+k-1}, u_{i+k}} \right) \\
	& & + \sum_{k=n/2 + 1}^{n-1} \sum_{i=0}^{n-1} \left( d_{u_i, u_{i-1}} + d_{u_{i-1}, u_{i-2}} + \cdots + d_{u_{i-n+k+1}, u_{i-n+k}} \right)\\
	& & + \sum_{i=0}^{n-1} \left( d_{u_i, u_{i+1}} + d_{u_{i+1}, u_{i+2}} + \cdots + d_{u_{i+n/2-1}, u_{i+n/2}} \right)\\
	&=& 2\left( 1+2+\cdots+(\frac{n}{2}-1) \right)\tau + \frac{n\tau}{2} 
	= \frac{n^2 \tau}{4}.
\end{eqnarray*}
% \begin{eqnarray*}
% 	\sum_{v \in V} \sum_{v' \in V \setminus \{v\}} d_{vv'} &=&
% 	\sum_{i=0}^{n-1} \sum_{k=1}^{n-1} d_{u_i, u_{i+k \, ({\rm mod} \, n)}} \\
% 	&\leq& 	\sum_{k=1}^{n/2} \sum_{i=0}^{n-1} h_{u_i, u_{i+k \, ({\rm mod} \, n)}}
% 	+ \sum_{k=n/2 + 1}^{n-1} \sum_{i=0}^{n-1} h_{u_i, u_{i+k \, ({\rm mod} \, n)}} \\
% 	&=& 2\left( 1+2+\cdots+(\frac{n}{2}-1) \right)\tau + \frac{n\tau}{2} 
% 	= \frac{n^2 \tau}{4}.
% \end{eqnarray*}

\noindent
(4) Since $v^*$ is a vertex 
	that attains $\min_{\, v \in V} \sum_{v' \in V \setminus\{v\}} d_{vv'}$,
	the inequality obtained in (3) directly implies the desired one.
\QED
}
\end{lemma}

Now we discuss the average of the traveling distances
	of $K^*_{\mathrm{DRR}}(m) \ (m \in \{0,1,\ldots,2n-3\})$. 
The traveling distance of a schedule is subject to the following constraint,
say the {\em athome\/} constraint:
each team stays at its home venue before a tournament 
and returns to its home venue after a tournament.
For simplicity of the analysis of the approximation ratio,
we temporary replace the athome constraint with the following assumption.

\medskip

\noindent
{\bf Assumption A.\@}\/
If a team plays away games at both the first and last slots,
	then the team moves from the venue of the last opponent to that of the first opponent,
	instead of the moves before the first slot and after the last slot.

\medskip

\begin{figure}
  \includegraphics[width=0.75\textwidth]{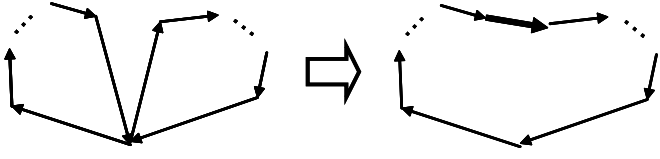}
\caption{Effect of Assumption A.}
\label{map}       % Give a unique label
\end{figure}

\noindent
We discuss a traveling distance of each team under Assumption~A\@.
Application of Assumption~A guarantees that 
	a route of each team in $K^*_{\mathrm{DRR}}(m) \ (m \in \{0,1,\ldots,2n-3\})$ 
	is a Hamilton cycle on~$G$ (see Figure~\ref{map}),
	and the traveling distance of~$K^*_{\mathrm{DRR}}(m)$ 
	is invariant with respect to $m \in \{0,1,\ldots,2n-3\}$. 
Thus, we only need to consider $K^*_{\mathrm{DRR}}$.
This assumption makes the analysis of the approximation ratio much easier.

Let the length of the cycle
	$(v_{0},v_{1},\ldots,v_{n-2})$ obtained by Christofides' method 
	in the proposed algorithm be~$\tau'$.
Note that $\tau' \leq (3/2) \tau$, 
	where $\tau$ denotes the length of a shortest Hamilton cycle on~$G$.
Analyzing the structure of $K^*_{\mathrm{DRR}}$ reveals the following lemma.
		
\begin{lemma}\label{lem3}
Under Assumption A, 
	the traveling distance of team $t$ in~$K^*_{\mathrm{DRR}}$ is bounded by
\[
	\left\{ \begin{array}{ll}
	\tau' + d_{v_t, v^*} + d_{v^*,v_{t+1}} - d_{v_t, v_{t+1}} 
	&  \quad (t \in \{0, 1,\ldots, n/2-1 \}), \\
	\tau' + d_{v_{t-1}, v^*} + d_{v^*,v_t} - d_{v_{t-1}, v_t}
	&  \quad (t \in \{n/2, n/2+1, \ldots, n-2\}), \\
	n\tau /2			&  \quad (t = n-1).
			\end{array} \right.
\]

{\rm \noindent
{\bf Proof.\/}
When $t \in \{0, 1, 2,\ldots, n/2-1 \}$, 
	team $t$ moves along a Hamilton cycle
	$(v_t,v^*,v_{t+1},\ldots,v_{n-2},v_0,v_1,v_2,\ldots,v_{t-1})$.
Consequently,
	the length of the tour is $\tau' + d_{v_t, v^*} + d_{v^*,v_{t+1}} - d_{v_t, v_{t+1}}$. 

When $t \in \{n/2, n/2+1, \ldots, n-2\}$, 
	a tour of team $t$ is a Hamilton cycle
	$(v_t,v_{t+1},\ldots,v_{n-2},v_0,v_1,v_2,\ldots,v_{t-1},v^*)$,
	and thus the length is $\tau' + d_{v_{t-1}, v^*} + d_{v^*,v_t} - d_{v_{t-1}, v_t}$. 

Since a tour of team $n-1$ is Hamiltonian,
	Lemma~\ref{lem2}(2) implies the desired result.
\QED
}
\end{lemma}

The above lemma implies an upper bound of the traveling distance of~$K^*_{\mathrm{DRR}}$.

\begin{lemma}\label{A-Distance}
Under Assumption~A, the traveling distance of~$K^*_{\mathrm{DRR}}$ 
	is bounded by
	$(n-2)\tau' + 2\sum_{v \in V \setminus \{v^*\}}d_{vv^*}+  (3/2)\tau + n\tau/2$.

{\rm \noindent
{\bf Proof.\/} 
Consider the sum total of upper bounds obtained in Lemma~\ref{lem3}
\[
	(n-1)\tau' + L + n\tau/2 
\]
where 
\begin{eqnarray*}
	L &\define&
		\sum_{t \in \{0, 1,\ldots, n/2-1 \}} 
		\left( d_{v_t, v^*} + d_{v^*,v_{t+1}} - d_{v_t, v_{t+1}} \right) \\
	&&	+ 	\sum_{t \in \{ n/2, n/2+1, \ldots, n-2 \}} 
		\left( d_{v_{t-1}, v^*} + d_{v^*,v_t} - d_{v_{t-1}, v_t} \right).
\end{eqnarray*}

\noindent
It is easy to see that 
\begin{eqnarray*}	
	L 
	&=&   \left( \sum_{t \in \{0, 1,\ldots, n/2-1 \}} d_{v_t, v^*} \right) 
		+ \left( \sum_{t \in \{1, 2,\ldots, n/2 \}} d_{v_t, v^*}   \right) \\
	&&	- \left( \sum_{t \in \{0, 1,\ldots, n/2-1 \}} d_{v_t, v_{t+1}} \right) 
		+ \left( \sum_{t \in \{ n/2 -1, n/2, \ldots, n-3 \}} d_{v_t, v^*} \right) \\
	&&	+ \left( \sum_{t \in \{ n/2, n/2+1, \ldots, n-2 \}} d_{v_t, v^*} \right)
		- \left( \sum_{t \in \{ n/2-1, n/2, \ldots, n-3 \}} d_{v_t, v_{t+1}} \right) \\
	&\leq&  2\sum_{v \in V \setminus \{v^*\}}d_{vv^*} - \sum_{t \in \{0, 1,\ldots, n-2 \}} d_{v_t, v_{t+1}} 
		+ d_{v_{n/2-1}, v^*} + d_{v_{n/2}, v^*} + d_{v_{n-2}, v_0} \\
	&\leq& 2\sum_{v \in V \setminus \{v^*\}}d_{vv^*} - \tau' + (3/2)\tau
\end{eqnarray*}
	where the last inequality follows from Lemma~\ref{lem2}(1).
From the above, the lemma holds.
\QED
}
\end{lemma}

Here we drop Assumption~A and restore the athome constraint, and consider the increase of the 
	traveling distance in the following lemma.

\begin{lemma}
For each team~$t$, 
	let $\ell_{\rm A} (t)$ be the traveling distance of~$t$
	in~$K^*_{\mathrm{DRR}}$ under Assumption~A\@.
Then, with the athome constraint the average of the traveling distances of team~$t$ 
	among $K^*_{\mathrm{DRR}}(m) \ (m \in \{0,1,\ldots,2n-3\})$ 
	is bounded by 
	$\ell_{\rm A} (t) + \sum_{v' \in V \setminus \{v\} }d_{vv'}/(n-1)$,
	where $v$ is the home venue of $t$.

{\rm \noindent
{\bf Proof.\/} 
For a choice $m \in \{0,1,\ldots,2n-3\}$, 
	every team $t'$ different from $t$ plays away game with~$t$ 
	at first slot just once.
Thus, the average length of the moves of team~$t$ 
	before the first slot is bounded by 
	$\sum_{v' \in V \setminus \{v\}}d_{vv'}/(2n-2)$.
Similarly, the average length of the moves of team~$t$ 
	after the last slot is bounded by 
	$\sum_{v' \in V \setminus \{v\}}d_{vv'}/(2n-2)$.
Thus, the average of the traveling distances of team~$t$
	is bounded by 
	$\ell_{\rm A} (t) +  \sum_{v' \in V \setminus \{v\}}d_{vv'}/(n-1)$.
\QED
}
\end{lemma}

Summarizing the above lemmas, we have the following theorem.

\begin{theorem}\label{thB}
The average of the total traveling distances 
	of schedules~$K^*_{\mathrm{DRR}}(m)$ $(m \in \{0,1,\ldots,2n-3\})$ 
	is bounded by 
\[
	(n-2)\tau' + 2\sum_{v \in V \setminus \{v^* \}}d_{vv^*}+  
	(3/2)\tau + n\tau/2 
	+ \sum_{v \in V} \sum_{v' \in V \setminus \{v\}} d_{vv'}/(n-1).
\]
\end{theorem}

Lastly we show the approximation ratio of the proposed algorithm.

\begin{theorem}\label{thm1}
The proposed algorithm is a\/ $2.75$-approximation algorithm for the UTTP\@.\\
{\rm \noindent
{\bf Proof.\/}
Let $z^*$ be the average of the total traveling distances of 
	schedules $K^*_{\mathrm{DRR}}(m)$ $(m \in \{0,1,\ldots , 2n-3\})$.
From Theorem~\ref{thB} and Lemma~\ref{lem2}(3)(4), we have:
\begin{eqnarray*}
z^*
%------------------------------------
&\leq& 
	(n-2)\tau' + 2\sum_{v \in V \setminus \{v^*\}}d_{vv^*}+  
	(3/2)\tau + n\tau/2 
	+ \sum_{v \in V} \sum_{v' \in V \setminus \{v\}} d_{vv'}/(n-1) \\
%------------------------------------
&\leq& (n-2)(3/2)\tau       + 2 n\tau /4 + (3/2)\tau + n\tau /2 +  (n^2 \tau /4)/(n-1) \\
%------------------------------------
&=&  (3/2)n \tau -3\tau +(1/2) n\tau + (3/2)\tau + (1/2)n\tau + (1/4)n\tau + (1/4)n\tau/(n-1) \\
%------------------------------------
&=&  (11/4) n\tau -(3/2)\tau + (1/4)n\tau/(n-1) \leq (11/4) n\tau. 
\end{eqnarray*}
The proposed algorithm output a best of
$K^*_{\mathrm{DRR}}(m)$ $(m \in \{0,1,\ldots , 2n-3\})$,
and thus the traveling distance of the output is at most~$z^*$. 
Since $n\tau$ is a lower bound of the distance of any double round-robin schedule, this concludes the proof. \QED
}
%%%%%%%%%%%%%%%%%%%
\end{theorem}

Let us consider a case that
	we have a shortest Hamilton cycle $H$ on $G$.
In this situation, the following corollary holds. 
\begin{corollary}\label{thm3}
If a shortest Hamilton cycle $H$ on $G$ is given, 
	there exists a\/ $2.25$-approximation algorithm for the UTTP\@.\\
{\rm \noindent
{\bf Proof.\/}
We replace a cycle obtained by Christofides' method in the proposed
	algorithm with a cycle obtained from $H$ by skipping vertex $v^*$.
Theorem~\ref{thB} implies that 
	the average of total traveling distances
	of schedules, say $z^{**}$, obtained by the proposed algorithm is bounded by
\begin{eqnarray*}
  z^{**} &\leq& 
	(n-2)\tau + 2\sum_{v \in V \setminus \{v^*\}}d_{vv^*}+  
	(3/2)\tau + n\tau/2 + \sum_{v \in V} \sum_{v' \in V \setminus \{v \}} d_{vv'}/(n-1) \\
	&\leq& n\tau -2\tau + 2 n\tau /4 + (3/2)\tau + n\tau /2 + (1/4)n\tau + (1/4)n\tau/(n-1) \\
	&=&  (9/4)n\tau -\tau /2 + (1/4)n\tau/(n-1) \leq (9/4)n\tau. 
\end{eqnarray*}
Thus, the approximation ratio is bounded by 2.25 in this case. \QED
}
\end{corollary}

% \section{Discussion}

% In this paper,
% 	we proposed a 2.75-approximation algorithm for the UTTP\@.
% The time complexity of the algorithm is bounded by $O(n^3)$, 
% 	because Christofides' algorithm requires $O(n^3)$ time.

% Let us consider a case that
% 	we have a shortest Hamilton cycle $H$ on $G$.
% We can replace a cycle obtained by Christofides' method in the proposed
% 	algorithm with a cycle obtained from $H$ by skipping vertex $v^*$.
% Theorem~\ref{thB} implies that 
% 	the average of total traveling distances
% 	of schedules, say $z^{**}$, obtained by the proposed algorithm is bounded by
% \begin{eqnarray*}
%   z^{**} &\leq& 
% 	(n-2)\tau + 2\sum_{v \in V \setminus \{v^*\}}d_{vv^*}+  
% 	(3/2)\tau + n\tau/2 + \sum_{v \in V} \sum_{v' \in V \setminus \{v \}} d_{vv'}/(n-1) \\
% 	&\leq& n\tau -2\tau + 2 n\tau /4 + (3/2)\tau + n\tau /2 + (1/4)n\tau + (1/4)n\tau/(n-1) \\
% 	&=&  (9/4)n\tau -\tau /2 + (1/4)n\tau/(n-1) \leq (9/4)n\tau. 
% \end{eqnarray*}
% Thus, the approximation ratio is bounded by 2.25 in this case.

\section{Computational Results}

In this section, we describe the results of computational experiments using the proposed approximation algorithm.

For the experiments, we took the distance matrices of NL and galaxy instances from the website~\cite{Trick_web}, because they are the most popular instances and one having the largest distance matrix (up to 40~teams), respectively.
We ran the proposed algorithm for the UTTP version of these instances;
to find a short Hamilton cycle, we use Concorde TSP solver~\cite{Concorde}.
It took less than one second to obtain a shortest Hamilton cycle even for the largest case ($n=40$).

To evaluate the quality of obtained solutions, we also tried to find optimal solutions of UTTP instances with integer programming.
Computations using integer programming were performed on the following PC:
Intel Xeon 3.33GHz$*2$, 24GB RAM, Windows~7 64bit, and Gurobi Optimizer 4.5.1~\cite{Gurobi} with 16 threads as an integer programming solver.
For both NL and galaxy instances:
for $n=4,6,8$ optimal solutions were obtained;
for $n=10$, after 500,000 seconds of computations, branch-and-bound procedures were not terminated;
for $n=12$ and larger instances, using integer programming it was difficult to find solutions better than those obtained by the proposed algorithm.

Tables~\ref{NL} and~\ref{galaxy} show the results of experiments for NL and galaxy instances, respectively.
The first columns of tables denote the number of teams,~$n$.
The second ones are the total traveling distance obtained by the proposed algorithm.
The third ones are the value of $n$ times the distance of a shortest Hamilton cycle, as a simple lower bound.
The fourth ones are the percentages of the gap between the second and third columns.

\begin{table}
\caption{Results for the UTTP version of NL instances}
\label{NL}       % Give a unique label
\begin{tabular}{crrrr}
\hline\noalign{\smallskip}
$n$ & approx. &$n*\mbox{TSP}$ & gap (\%)${}^\dagger$& best UB  \\
\noalign{\smallskip}\hline\noalign{\smallskip}
 4 &       8276 &   8044 &  2.9 &  8276${}^\ddagger$ \\
 6 &      20547 &  17826 & 15.3 & 19900${}^\ddagger$ \\
 8 &      33190 &  27840 & 19.2 & 30700${}^\ddagger$ \\
10 &      47930 &  38340 & 25.0 & 45605${}^\star$ \\
12 &      81712 &  67200 & 21.6 &      \\
14 &     128358 & 103978 & 23.4 &      \\
16 &     156828 & 119088 & 31.7 &      \\
\noalign{\smallskip}\hline \noalign{\smallskip}
\multicolumn{5}{l}{${}^\dagger$gap is obtained by $(\frac{\mathrm{approx.}}{n*\mathrm{TSP}}-1)*100.0$} \\
\multicolumn{5}{l}{${}^\ddagger$optimal}\\
\multicolumn{5}{l}{${}^\star$best incumbent solution after 500,000 seconds}
\end{tabular}
\end{table}

\begin{table}
\caption{Results for the UTTP version of galaxy instances}
\label{galaxy}       % Give a unique label
\begin{tabular}{crrrr}
\hline\noalign{\smallskip}
$n$ & approx. &$n*\mbox{TSP}$ & gap (\%)${}^\dagger$& best UB  \\
\noalign{\smallskip}\hline\noalign{\smallskip}
 4 &    416 &    412 &  1.0  & 416${}^\ddagger$   \\
 6 &   1197 &   1068 & 12.1  & 1178${}^\ddagger$  \\
 8 &   2076 &   1672 & 24.2  & 1890${}^\ddagger$  \\
10 &   3676 &   3020 & 21.7  & 3570${}^\star$  \\
12 &   5514 &   4524 & 21.9  & \\
14 &   7611 &   6216 & 22.4  & \\
16 &   9295 &   7408 & 25.5  & \\
18 &  12320 &  10026 & 22.9  & \\
20 &  14739 &  11880 & 24.1  & \\
22 &  19525 &  16522 & 18.2  & \\
24 &  25026 &  21216 & 18.0  & \\
26 &  32250 &  27846 & 15.8  & \\
28 &  41843 &  36708 & 14.0  & \\
30 &  52073 &  46410 & 12.2  & \\
32 &  62093 &  55104 & 12.7  & \\
34 &  77392 &  69326 & 11.6  & \\
36 &  88721 &  78624 & 12.8  & \\
38 & 103988 &  92568 & 12.3  & \\
40 & 120895 & 107800 & 12.1  & \\
\noalign{\smallskip}\hline \noalign{\smallskip}
\multicolumn{5}{l}{${}^\dagger$gap is obtained by $(\frac{\mathrm{approx.}}{n*\mathrm{TSP}}-1)*100.0$} \\
\multicolumn{5}{l}{${}^\ddagger$optimal}\\
\multicolumn{5}{l}{${}^\star$best incumbent solution after 500,000 seconds}
\end{tabular}
\end{table}

Like most theoretical approximation algorithms, the obtained gaps are much better than the theoretical approximation ratio 2.75 (175\% gap).
For the NL instances and the galaxy instances of up to 20~teams, the gap is around~25\%.
For the galaxy instances of more than 20~teams, the gap is less than 20\%.
Note that the gaps shown in the tables are from the ratio of the obtained distance to a lower bound, but not to optimal distance.
Therefore the gaps between the obtained distance and the optimal value are still better than the gaps shown in the tables.

\section{Conclusion}
This paper proposed an approximation algorithm for the unconstrained traveling tournament problem, 
	which is a variant of the traveling tournament problem. 
The approximation ratio of the proposed algorithm is 2.75, 
	and the algorithm yields a solution satisfying the no-repeater and mirrored constraints.
If a shortest Hamilton cycle on the home venues of the teams is available, the approximation ratio is improved to 2.25. 
Computational experiments showed that the algorithm generates solutions of good quality; 
	the gap between the obtained solution and a simple lower bound is around~25\% for small instances (up to 20~teams)
	and is less than 20\% for larger instances.

%%%%%%%%%%%%%%%%%%

%\begin{acknowledgements}
%\end{acknowledgements}


\end{document}